\documentclass[onecolumn,showpacs,preprintnumbers,amsmath,amssymb]{revtex4}
\usepackage{graphicx}
\usepackage{dcolumn}
\usepackage{bm}

\newcommand{\e}{{\mathrm e}}

\newcommand{\ee}{\e^+\e^-}

\begin{document}

\preprint{hep-ph/0507051}

\title{ Micro-canonical pentaquark production in $\ee$ annihilations }

\author{Fu-Ming Liu}
\email{liufm@iopp.ccnu.edu.cn}
\affiliation{
Institute of Particle Physics, Central China Normal University, Wuhan, China
}%

\author{Klaus Werner}
\affiliation{
Laboratoire SUBATECH, University of Nantes - IN2P3/CNRS
- Ecole des Mines de Nantes, Nantes, France
}%

\date{\today}

\begin{abstract}
The existence of pentaquarks, namely baryonic states made up of four quarks 
and one antiquark, became questionable, because 
the candidates, i.e. the $\Theta^+$ peak, are seen in certain reactions, 
i.e. p+p collisions, but not in others, i.e. $\ee$ annihilations.
In this paper, we estimate the production of $\Theta ^{+}(1540)$ and
$\Xi^{--} (1860)$ in $\ee$ annihilations at different energies 
using Fermi statistical model as originally proposed in its microcanonical form.
The results is compared with that from pp collisions at SPS and RHIC energies. 
We find that, if pentaquark states exist, the production is highly 
possible in $\ee$ annihilations. For example, at LEP energy $\sqrt{s}$=91.2~GeV, both $\Theta ^{+}(1540)$ and $\Xi^{--} (1860)$ yield more than in pp collisions 
at SPS and RHIC energy.
\end{abstract}


\maketitle

\section{Introduction}
Pentaquark is the name for baryons made up of four quarks and one
antiquark. Recent advances in theoretical and experimental work
led to the observation of pentaquark candidates by some
experiments, however, other experiments reported no observation
from their search. For more details, see the review of the
experimental evidence on pentaquarks and critical discussion
\cite{kabana}. The question of the existence of pentaquarks is raised again,
due to the non-observation in so many experiments. To answer the
question, it is important to assess the cross sections in
different processes.

The candidate $\Theta^+$ peak is seen in different reactions namely of
$\gamma$+A, $\mu$+A, p+p, K+Xe, e+d, e+p, K+Xe\cite{positive}\cite{NA49}.
All of those reactions involved at least one baryon in the initial state.
But other experiments, e.g.
 $e^+e^-$: Babar, Belle, Bes, LEP experiments\cite{Armstrong},
$p \overline{p}$: CDF, D0
 $pA$:E690,
$\gamma p$: FOCUS,
 $pA$: HERA-B,
 $ep$: Zeus (for the $\theta^0_c$)
$ \mu^+ \ ^6 Li D$: COMPASS,
Hadronic Z decays: LEP,
$\pi$, $K$, p on A: HyperCP,
$\gamma \gamma$: L3,
$\pi, p, \Sigma$ on p: SELEX,
$pA$: SPHINX,
$\Sigma^- A$: WA89
$K^+ p$: LASS\cite{non},
 did not observe those candidates.
Because the experiments such as $e^+e^-$ did not observe those candidates,
it is generally believed a non-zero initial baryon number is 
essential to the pentaquark production, i.e. pp collisions or collisions involved with nuclei, as we can see the collision types of the positive reports.
To check this, we calculate in this paper the pentaquark 
production in $\ee$ annihilations.
A theoretical comparison of pentaquark production between p+p\cite{penta-pp} and  $\ee$
processes may be helpful to the question if pentaquarks exist.

Why we choose the microcanonical approach to calculate?
Traditionally, the hadron production in $\ee$ scattering at high
energy is treated as a two-stage process. First
$\ee \rightarrow \gamma^{\star}/Z \rightarrow q\bar{q} $
is calculated using perturbative theory.
 Then the $q\bar{q}$ system  produces hadrons phenomenologically based on 
string fragmentation or cluster fragmentation.
Pentaquark states are exotic, hard to treated within the frame of conventional 
string models. 
In this situation statistical approaches may be of great 
help\cite{Fermi:1950jd},\cite{Landau}.
It was Hagedorn who introduced statistical methods into
the strong interaction physics in order to calculate the momentum
spectra of the produced particles and the production of strange particles\cite{hag1}.
Later, after statistical models have been successfully applied to
relativistic heavy ion collisions \cite{Hagedorn}, 
Becattini and Heinz \cite{bec} came back to the statistical description
of elementary reactions.

According to the situations of obeying conservation laws, statistical models are classified into four ensembles:

{*} microcanonical: both, material conservation laws (Q, B, S, C,
\( \cdots  \) ) and motional conservation laws (E, \( \overrightarrow{p} \),
\( \overrightarrow{J} \), \( \cdots  \)), hold exactly.

{*} canonical: material conservation laws hold exactly, but motional
conservation laws hold on average (a temperature is introduced).

{*} grand-canonical: both material conservation laws and motional
conservation laws hold on the average (temperature and chemical potentials
introduced).

{*} mixed ensemble: for example, to estimate particle production in heavy ion collisions, so-called partially canonical and partially grand-canonical models are employed where strangeness conserves strictly, but the conservation of net charge hold on the average.

What one expects is that the microcanonical ensemble
must be used for very small volumes, i.e. the systems created by the collisions between elementary particles. For intermediate volumes the
canonical ensemble should be a good approximation, while for very
large volumes, i.e. the systems created by heavy ion collisions,
 the grand-canonical ensemble can be employed. 
Therefore we take a microcanonical approach to calculate the 
production of pentaquark states from electron-positron annihilations.

\section{The approach}

The idea was originally proposed by Fermi in its microcanonical form\cite{Fermi:1950jd} and realized with Markov chain technique\cite{wer}.
We calculate the hadron production in $\ee$ annihilations at a given energy
as a statistical decay of a cluster which carries net quark contents
 \( Q=(N_{u}-N_{\bar{u}},N_{d}-N_{\bar{d}},N_{s}-N_{\bar{s}})=(0,0,0) \).
The cluster is charactered by three parameters,  
cluster energy (mass) $E$, 
 volume $V$ and strangeness suppression factor $\gamma_s$.
We assume that hadron production from the cluster 
is dominated by the  n-body phase space.
More precisely, the probability of the cluster hadronization into a
 configuration \( K=\{h_{1},p_{1};\ldots ;h_{n},p_{n}\} \)
of hadrons \( h_{i} \) with four momenta \( p_{i} \) is given by
the microcanonical partition function \( \Omega (K) \) of an ideal,
relativistic gas of the \( n \) hadrons \cite{wer},
\begin{eqnarray}
{\Omega}^{(\text{K})}=&&\frac{V^{n}}{(2\pi \hbar )^{3n}}\, \prod _{i=1}^{n}{g_{i}\gamma_s ^{s_i}} \,
\prod _{\alpha \in \mathcal{S}}\, \frac{1}{n_{\alpha }!}\,
\prod _{i=1}^{n}d^{3}p_{i}\, 
\delta (E-\Sigma \varepsilon _{i})\, \delta (\Sigma \vec{p}_{i})\, \delta _{Q,\Sigma q_{i}},
\label{eq:omega}
\end{eqnarray}
 with \( \varepsilon _{i}=\sqrt{m_{i}^{2}+p_{i}^{2}} \) being the
energy, and \( \vec{p}_{i} \) the 3-momentum of particle \( i \).

The term \( \delta _{Q,\Sigma q_{i}} \) ensures flavour conservation;
\( q_{i} \) is the flavour vector of hadron \( i \). The symbol
\( \mathcal{S} \) represents the set of hadron species considered:
we take \( \mathcal{S} \) to contain the pseudoscalar and vector
mesons \( (\pi ,K,\eta ,\eta ',\rho ,K^{*},\omega ,\phi ) \) and
the lowest spin-\( \frac{1}{2} \) and spin-\( \frac{3}{2} \) baryons
\( (N,\Lambda ,\Sigma ,\Xi ,\Delta ,\Sigma ^{*},\Xi ^{*},\Omega ) \)
and the corresponding antibaryons. \( n_{\alpha } \) is the number
of hadrons of species \( \alpha  \), and \( g_{i} \) is the degeneracy
of particle \( i \).

It's well known that strangeness will be overpopulated 
if the hadron production is purely determined by the $n$-body phase space.
The common treatment is to introduce so-called strangeness suppression factor 
$\gamma_s$ with $ 0< \gamma_s <1 $. So we employ this factor as well in Eq. (1), 
and the index $s_i$ is the number of (anti)strange components in the 
final-state particle $i$, i.e. 
for Kaons, $\Lambda$, $\Sigma$ and their antibaryons, $s_i=1$;
for $\phi$-meson, $\Xi$ and $\bar\Xi$, $s_i=2$;
for  $\Omega$ and $\bar\Omega$, $s_i=3$.

Similar to the previous work\cite{penta-pp}, we add the pentaquark states
$\Theta ^{+}(1540)$ and $\Xi (1860)$  into $\mathcal{S}$.
The $\Theta ^{+}$ has quark
contents $[uudd\bar{s}]$. The $\Xi (1860)$ can be $\Xi ^{--}[ddss\bar{u}]$,
$\Xi ^{-}[dssu\bar{u}]$ , $\Xi ^{0}[ussd\bar{d}]$ or $\Xi ^{+}[uuss\bar{d}]$.
The spin of pentaquark states can not be determined by experiments
yet, and it is generally accepted pentaquark states are spin-$\frac{1}{2}$ particles,
so we take a degeneracy factor $g=2$. 
As for strangeness suppression, $s_i=1$ for $\Theta^+$  
and  $s_i=2$ for $\Xi$(1860).

The high dimensional phase space integral is verified via constructing Markov chains of hadron configurations $K$. The Metropolis algorithm provides random configurations $K$ according to the weight, the corresponding microcanonical partition function $\Omega(K)$. All possible random configurations are included. 

Working with Markov chains one has to worry about two kinds of convergences: the
number of iterations per chain must be sufficiently big (essentially a multiple of the so-called auto-correlation time), otherwise the method is simply wrong. 
Secondly, the number of simulated chains must be sufficient large, to obtain the desired statistical accuracy. Questions related to the auto-correlation time have been studied in detail in earlier publications\cite{wer}, so that the error due to auto-correlations can be neglected. The statistical error will be treated carefully at the result section. 

In addition to checking auto-correlations and statistical errors,
also physics cross checks have been performed: a comparison 
\cite{FM} of our Monte Carlo method with canonical method 
gives a good agreement when the systems have big volumes, i.e. 50fm$^3$ 
and high energies, i.e. 10 times the mass of observed particles.

We generate randomly configurations \( K \) according to the
probability distribution \( \Omega (K) \). The Monte Carlo
technique allows to calculate mean values of an observable as
\begin{eqnarray}
\bar{A}=\sum _{K}A(K)\, \Omega (K)/\sum _{K'}\Omega (K'),
\label{eq:bbb}
\end{eqnarray}
 where \( \sum  \) means summation over all possible configurations
and integration over the \( p_{i} \) variables. \( A(K) \) is some
observable assigned to each configuration, as for example the $4\pi$ 
multiplicity
\( M_{h}(K) \) of hadrons of species \( h \) present in \( K \).
Since \( \bar{M}_{h} \) depends on \( E \), \( V \) and \( \gamma_s \), 
we usually write \( \bar{M}_{h}(E,V, \gamma_s) \).

The cluster parameters energy  $E$, volume $V$ and strange factor $\gamma_s$
are chosen so that the cluster decay reproduces best the measured multiplicity
of the selected hadrons in $\ee$ annihilations at a given energy of 
\( \sqrt{s} \). 
This is achieved by minimizing \( \chi ^{2} \): 
\begin{eqnarray}
\chi ^{2}(E,V,\gamma_s)=\sum _{j=1}^{\alpha }\frac{[\bar{M}_{\mathrm{exp},j}(\sqrt{s})-\bar{M}_{j}(E,V,\gamma_s)]^{2}}{\sigma _{j}^{2}}
\label{eq:ccc}
\end{eqnarray}
 where \( \bar{M}_{\mathrm{exp},j}(\sqrt{s}) \), and \( \sigma _{j} \)
are the experimentally measured multiplicity and its error of some selected hadron species \( j \) in $\ee$ collisions at an energy of \( \sqrt{s} \).

With the cluster parameters determined by the selected hadrons, we can predict
the multiplicity of any particles included in the hadron species list, i.e. 
pentaquark states from $\ee$ collisions at the energy of \( \sqrt{s} \).  

\section{The Results}
We calculate pentaquark production from  $\ee$ collisions at \( \sqrt{s} =29, 35, 91.2\)~GeV. This is realized by two steps: first, we have to determine the 
cluster parameters energy  $E$, volume $V$ and strange factor $\gamma_s$ for each 
collision energy \( \sqrt{s}\); then, we use the obtained cluster parameters to 
calculate pentaquark production.

\begin{figure}
\includegraphics[scale=.75]{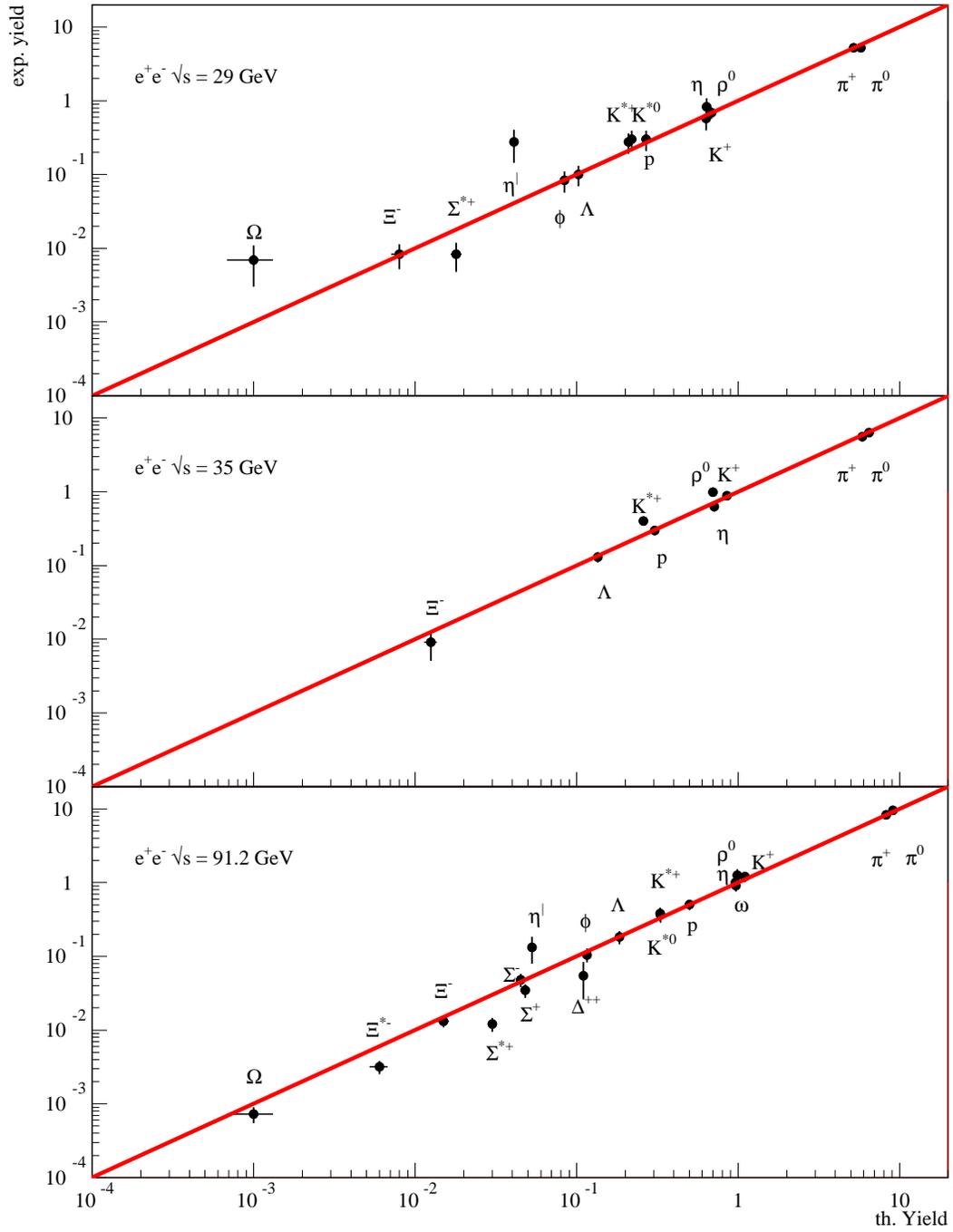}
\caption{\label{fig:eefit3p}(Color online)
A comparison between hadron production from cluster decay and from 
\(\ee\) annihilation experiments at 29~GeV, 35~GeV, and 91.2~GeV. }
\end{figure}

For the first step: with data taken from \cite{exp_ee}, we select \( j=\pi ^{+}, p, K^+, \Lambda \) and their antiparticles to determine the  cluster parameters energy  $E$, volume $V$ and strange factor $\gamma_s$. The available experimental yields of other particles are used to check the reliability of our results.
For each set of parameters ($E$, $V$, $\gamma_s$), the multiplicities of selected hadrons are calculated from 10,000 random hadron configurations so that the relative statistical errors of the selected hadrons are within 0.03.  Then the  best-chosen set of parameters is obtained by minimizing \( \chi ^{2} \) as in Eq.(3).
Fig.\ref{fig:eefit3p} displays the results of our fit procedure 
in comparison with the experimental data at different energies.
Here the error bars of theoretical yields are statistical.

Fig.1 tells that the best-chosen cluster parameters with the experimental multiplicities of \( \pi ^{+}, p, K^+, \Lambda \) from $\ee$
collisions at the energy of $\sqrt{s}=29$~GeV
can also reproduce very well the yields of other particles such as 
 \( \pi^0, \rho^0, \eta, \phi, \Xi^- \).
The less well reproduced particles are \( \eta' \),  \( {\Sigma^{*}}^{+} \) and \( \Omega \), spin-1 meson and spin-$\frac{3}{2}$ baryons.
 
At the energy of $\sqrt{s}=35$~GeV, the best-chosen cluster parameters with 
the experimental yields of \( \pi ^{+}, p, K^+, \Lambda \)  
can also reproduce very well the yields of other particles such as 
 \( \pi ^{0}, \rho^0, \eta,  {K^{*}}^{+}, \Xi^- \).

At the energy of $\sqrt{s}=91.2$~GeV, the best-chosen cluster parameters with
the experimental yields of \( \pi ^{+}, p, K^+, \Lambda \) from $\ee$
collisions can well reproduce the yields of other particles
such as \(\pi ^{0}, \rho^0, \eta, \omega,  K^*, \phi, \Sigma, \Xi^- \).
The less well reproduced particles are \( \eta' \),  \( \Delta^{++}, 
\Sigma^*, \Xi^* \) and \( \Omega \), spin-1 meson and spin-$\frac{3}{2}$ baryons.

The large list of reproduced particle proves that the cluster parameters,
energy  $E$, volume $V$ and strange factor $\gamma_s$,
 determined by 4 particle yields of \( \pi ^{+}, p, K^+, \Lambda \)
can be reliably used to estimate the yields of particles, especially  
for spin-0 mesons and spin-$\frac{1}{2}$ baryons. 
As it is accepted that the spin of pentaquark states is 
 $\frac{1}{2}$, our estimation on pentaquark yields does not
 suffer from the problem with spin-3/2 baryon yields.

In Tab. \ref{tab:table1} the best-chosen set of parameters ($E$, $V$ $\gamma_s$), for $\ee$ collisions at the energy of $\sqrt{s}=$29, 35  and 91.2~GeV are collected. The small $\chi^2$/dof indicates a good fit quality. 

We find that with the increase of $\ee$ energy $\sqrt{s}$, the created cluster has a higher energy $E$ and a bigger volume $V$. The strange factor $\gamma_s$
does not change much with the increase of $\ee$ energy $\sqrt{s}$.

One may feel hard to understand why $E=\sqrt{s}$ does not hold, instead, 
the created cluster has an energy $E$ much smaller than the $\ee$ energy 
$\sqrt{s}$. 
Let's imagine an expanding fireball (mainly longitudinal/thrust) created in $\ee$ annihilation. In this case, there is a lot of collective kinetic energy.
The parameter $E$ means the sum of the energies of volume elements in their proper frames, in other words their invariant masses. So this effective mass is much smaller than the mass of the total system, which is of course  $\ee$ energy $\sqrt{s}$. If we only consider total yields, we do not need to specify the details of the collective expansion. The price is that this simple 3-parameter model cannot make any statements about transverse momentum spectra or rapidity spectra.

The volume we find here is much bigger than that in Fermi's model \cite{Fermi:1950jd}, where the volume is estimated according to the colliding particles with the Lorentz contact. In our case, the volume $V$ is the size of the collision system at the moment of hadronization. After $\ee$ annihilates, the collision system gets many quark pair production and expanding with time. So the size is much much bigger than that of an electron and a positron. 

The strangeness suppression factor $\gamma_s$ changes very little with the collision energy $\sqrt{s}$. The value is around 0.7, bigger than what we found in pp collisions. This is also consistent with the result from canonical fitting by Becattini\cite{bec}.

\begin{table}
\caption{\label{tab:table1}
Cluster parameters for $\ee$ annihilations at different energies
determined by the yields of 
\( \pi ^{+}, p, K^+, \Lambda \) and their antiparticles. }
\begin{ruledtabular}
\begin{tabular}{lllll}

    & $E(\mathrm{GeV})$   & $V(\mathrm{fm}^3)$   & $\gamma_s$ &   $\chi^2$/dof \\
\hline
$\sqrt{s}=$29~GeV    & 10.8$\pm$0.8  & 62$\pm$24 & 0.64$\pm$0.11  &  1.97/5  \\
$\sqrt{s}=$35~GeV    & 12.4$\pm$0.8  & 70$\pm$20 & 0.71$\pm$0.14  &  4.97/5  \\ 
$\sqrt{s}=$91.2~GeV  & 17.6$\pm$0.4  & 102$\pm$12 & 0.61$\pm$0.07 & 2.92/5 \\
\end{tabular}
\end{ruledtabular}
\end{table}

\begin{figure}
\includegraphics[scale=.75]{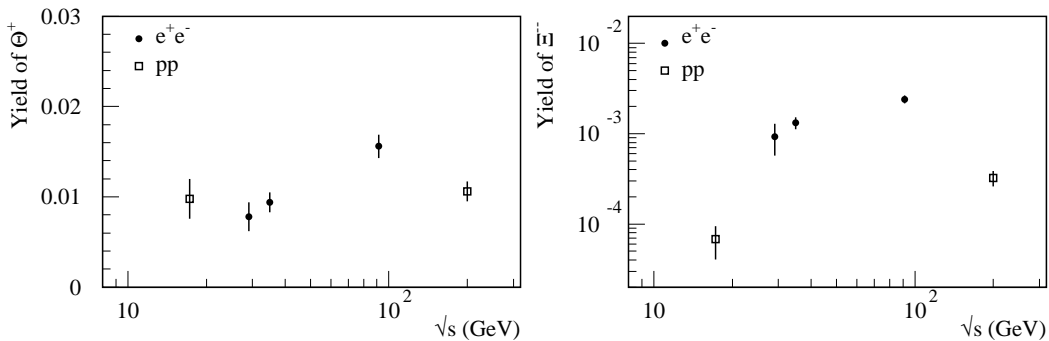}
\caption{\label{fig:1penta}
A comparison of the $4\pi$ yields of $\Theta^+$ (left) and $\Xi^{--}$(right) 
from \(\ee\) annihilations and from pp collisions at\cite{penta-pp}.  }
\end{figure}

With the cluster parameters obtained above, we can calculate the production
of all hadrons in the hadron list. 
Pentaquark states have been included in the hadron list, so we can get the $4\pi$
yields of pentaquark states from $\ee$ annihilations at the energy of 29, 35, and 91.2~GeV. Since the yields of pentaquark states are much lower than ordinary hadrons, we calculate their yields from the average of 1,000,000 random configurations with Eq.(2). 
In Fig. \ref{fig:1penta} we plot the result as  $4\pi$ yields of pentaquark 
states versus the center mass system energy $\sqrt{s}$, and compare with that 
from pp collisions at SPS energy E$_\mathrm{lab}$=158~GeV and RHIC energy E$_\mathrm{cms}$=200~GeV. The error bar is a combination of the statistical Monte Carlo evaluation error with the influence of the fit error of the parameters. The two errors are at the same magnitude, and the statistical one is smaller.
 
The production of pentaquark states, i.e. $\Theta^+$ and $\Xi^{--}$, from \(\ee\) annihilations increase with the collision energy.
In pp collisions, we can see the similar behavior for all pentaquark states except $\Theta^+$. 
The production of $\Theta^+$ in pp collisions is relatively high at a few GeV, decreases slightly, then increases at teens GeV and keep increasing for higher collision energies\cite{penta-pp}. 
The special behavior of $\Theta^+$ in pp collisions at low energies is due to the proton excitation, which does not apply in \(\ee\) collisions.

The yields of $\Theta^+$ from $\ee$ collisions at the energies we study are at the same magnitude of that from pp collisions. The $\Theta^+$ production from $\ee$ at LEP energy  $\sqrt{s}=$91.2~GeV is even higher than the yields from pp
collisions at both SPS and RHIC energy. 

For $\Xi^{--}$, we get obviously higher production from  $\ee$ collisions
than from pp collisions. This can be understood: \\
 1. The average mass of clusters created in  $\ee$ collisions is much bigger than 
that in pp collisions, i.e. the clusters have average mass 17.6~GeV for $\ee$ at
 $\sqrt{s}=$91.2~GeV, and mass 7.3~GeV for pp collisions at SPS and mass 16.15~GeV at RHIC energy. The cluster mass is most sensitive parameter for particle production. Certainly cluster with big mass can produce more particles. \\
2. The strangeness suppression factor is bigger in  $\ee$ collisions. $\Xi^{--}$ contains two strange quarks, and gets squared strangeness suppression, which is about 0.1 in pp collisions and about 0.4 in  $\ee$ collisions.

The particle ratios $\Theta^+ /p$ and  $\Xi ^{--}(1860)/\Xi ^{-}$ are 
of interest and have been discussed by several different approaches
in heavy collisions and pp collisions. Here we study the ratios in $\ee$
collisions at different energies. 


We get the $\Theta^+ /p $ about 0.03 in $\ee$ annihilations, while it is 
much smaller, about 0.007,  in pp collisions with the microcanonical 
approach\cite{penta-pp}. 
We recognize the big difference can be caused by the strangeness suppression factor, 0.6 $\sim $ 0.7 in $\ee$ annihilations, but 0.33 in pp collisions. 
The grandcanonical ensemble\cite{Randrup:2003fq} 
gives about 0.06 for this ratio in heavy ion collisions, which is even higher.
This occurs when the strange chemical potential $\mu_s=0$, which 
corresponds to $\gamma_s=1$, an even bigger strangeness suppression factor. 

We get higher $\Xi ^{--}(1860)/\Xi ^{-}$ ratio, about 0.1, in $\ee$ annihilations, which is 0.02 in pp collisions with the microcanonical approach\cite{penta-pp}, and 0.01 in heavy ion collisions obtained by the grandcanonical results\cite{Letessier:2003by}. 

\section{Discussion and Conclusion}

We estimate the pentaquark production, i.e. $\Theta^+$ and $\Xi^{--}$, from $\ee$ annihilations at the energy of $\sqrt{s}$=29, 35 and 91.2~GeV using Fermi statistical model as originally proposed in its microcanonical form.
We obtain increasing production of pentaquark states with the increase of collision energy. Comparing with the previous work for pp collisions, we find that the
yields of $\Theta^+$, from $\ee$ at the above mentioned energies are  
at the same  magnitude as those from pp collisions at SPS and RHIC energies.
The yields of $\Theta^+$  from $\ee$ at LEP energy $\sqrt{s}$=91.2~GeV
is higher than the yields from pp collisions at SPS and RHIC energies.

From our estimation, the production of $\Xi^{--}$ from  $\ee$ collisions is obviously higher than that from pp collisions at SPS and RHIC energies. That's very different from the experimental report -- the observation of  $\Xi^{--}$ was reported for the first time in the SPS experiment, by NA49 collaboration\cite{NA49}; but none of the four collaborations in LEP experiment get the observation, though they did search. 

Theoretically we conclude that Fermi statistical model pentaquark production
is quite high from $\ee$ annihilations if pentaquark states do exist due to a very high energy clusters are created when the $\ee$ annihilate. The energy parameter $E$=17.6~GeV for $\ee$ at $\sqrt{s}=$91.2~GeV, while $E$=16.2~GeV for pp collisions at $\sqrt{s}=$200~GeV.
The average proper mass of clusters created in pp collisions at RHIC energy is lower, indicates a lot of energy is taken away collectively, due to the existence of the leading particles.
Initial baryons are not necessary for pentaquark production.
The clusters with big masses created from any kind of
high energy collisions can provide a rather high yield.

\section{Acknowledgments}

This work was sponsored by National Natural Science Foundation of China (NSFC $^{\#}$10447110 and $^{\#}$10505010) and the Scientific Research Foundation for the Returned Overseas Chinese Scholars, State Education Ministry (SRF for ROCS, SEM, (2005)$^{\#}$383).
FML thanks Prof. B.Q. Ma for fruitful discussions.

\end{document}